\documentclass[12pt, oneside]{article}   	
\usepackage{geometry} 
\usepackage{amsmath} 								
\usepackage{graphicx}	
\usepackage{amssymb}
\usepackage{cite}
\usepackage{color,xspace}
\usepackage{extarrows}
\usepackage{caption}

\numberwithin{equation}{section}
\DeclareSymbolFont{extraup}{U}{zavm}{m}{n}
\DeclareMathSymbol{\vardiamond}{\mathalpha}{extraup}{87}
\usepackage{float}
\restylefloat{table}
\allowdisplaybreaks

\def\twomat[#1,#2][#3,#4]{\left( \begin{array}{cc} #1 & #2 \\ #3 & #4 \end{array} \right)}
\def\thv[#1,#2,#3]{\left( \begin{array}{c} #1 \\ #2 \\ #3 \end{array} \right)}
\def\twv[#1,#2]{\left( \begin{array}{c} #1 \\ #2 \end{array} \right)}

\title{Notes on }

\date{}

\begin{document}

\begin{flushright}
\end{flushright}
\begin{center}

\vspace{1cm}
{\LARGE{\bf Weak Gravity Conjecture in de Sitter space-time}}

\vspace{1cm}

\large{\bf Ignatios Antoniadis$^{a,b,\vardiamond}$ \let\thefootnote\relax\footnote{$^\vardiamond$antoniad@lpthe.jussieu.fr}
and Karim Benakli$^{a,\spadesuit}$ \let\thefootnote\relax\footnote{$^\spadesuit$kbenakli@lpthe.jussieu.fr}
 \\[5mm]}

{ \sl $^{a}$Laboratoire de Physique Th\'eorique et Hautes Energies (LPTHE),\\ UMR 7589,
Sorbonne Universit\'e et CNRS, \\ 4 place Jussieu, 75252 Paris Cedex 05, France.}

{\sl $^{b}$Albert Einstein Center, Institute for Theoretical Physics, University of Bern, \\
 			Sidlerstrasse 5, 3012 Bern, Switzerland.}

\end{center}
\vspace{0.7cm}

\abstract{
We propose a generalisation of the Weak Gravity Conjecture in de Sitter space by studying charged black-holes and comparing the gravitational and an abelian gauge forces. 
Using the same condition as in flat space, namely the absence of black-hole remnants, one finds that for a given mass $m$ there should be a state with a charge $q$ bigger than a minimal value $q_{\rm min}(m,l)$, depending on the mass and the de Sitter radius $l$, in Planck units. In the large radius flat space limit (large $l$), $q_{\rm min}\to m$ leading to the known result $q>m/\sqrt{2}$, while in the highly curved case (small $l$) $q_{\rm min}$ behaves as $\sqrt{ml}$.
We also discuss the example of the gauged R-symmetry in $N=1$ supergravity.
}

\newpage
\setcounter{footnote}{0}

\section{Introduction}
\label{introduction}


Cosmological observations favour the presence of a non-vanishing positive vacuum energy with an equation of state consistent with a cosmological constant. On the other hand, the difficulties in constructing de Sitter solutions in string theory (see for example~\cite{Maldacena:2000mw,Danielsson:2018ztv}) have motivated the de Sitter conjecture stating that such a space-time is absent in a consistent theory of gravity~\cite{Obied:2018sgi}. 
However, this conjecture is based mainly on classical arguments which seem to be invalidated at the string quantum level~\cite{Antoniadis:2018hqy}. 
Moreover, on the basis of our need to describe the observable Universe, de Sitter space should arise at least as an "approximate" background, of course in a yet unknown sense, of some consistent string theory vacua.
The present work should be taken in this sense.

The Weak Gravity Conjecture (WGC)~\cite{ArkaniHamed:2006dz} is probably the best established among the swampland conjectures~\cite{Vafa:2005ui}.  It states that for a $U(1)$ gauge symmetry with gauge coupling $g$, there is a least one state that has a charge $q$ bigger than its mass $m$, measured in Planck units, $8\pi G=\kappa^2=1$:
\begin{equation}
 m < \sqrt{2} g q \, .
\label{WGC}
\end{equation}
Black holes represent a unique playground for investigating many of the features of quantum gravity. For instance, the WGC appears to be related with the requirement that black holes should be able to decay without leaving remnants. Indeed, in the case of a black hole in asymptotically flat space-time, the arguments presented in~\cite{ArkaniHamed:2006dz} follow from the conservation of charge and energy in the black hole decay, which should then result in states with sum of masses smaller than the mass of the original black hole.

Our goal is to find how (\ref{WGC}) is modified in the presence of a positive vacuum energy. The problem of finding in de Sitter space the form of the WGC has already been discussed in the literature. In particular, Ref.~\cite{Montero:2019ekk} has suggested a possible relation between charges and masses obtained by studying  the decay of the Nariai \cite{Nariai} or the de Sitter Reissner-Nordstr\o m black holes  \cite{Romans:1991nq}. They required that the decay should proceed in a slow rate such that it allows to end up in a de Sitter space-time and not to tunnel into a big crunch. The derived constraint differs substantially from our finding as it does not lead to (\ref{WGC}) in a manifest way when the cosmological constant is taken to zero. An other argument for the WGC in de Sitter space, presented in \cite{Huang:2006hc}, is that a minimally charged monopole does not form a black hole. This will be reviewed in some details at the end of Section 4. The resulting constraint, as stated in \cite{Huang:2006hc} amounts to prohibiting global symmetries when the flat space-time limit is taken. But as this constraint does not involve the mass of the states, it does not connect either to (\ref{WGC}). One should keep in mind that we live in a universe with a non-vanishing vacuum energy while flat space time laws are in extremely good agreement with every day experiments. This indicates that there must be an expansion such that when the vacuum energy is taken to be negligible, flat space-time laws, such as (\ref{WGC}), should be recovered.

In order to derive the form of the weak gravity conjecture in de Sitter background (dS-WGC), we will consider here another facet of the black hole physics based on arguments that led to (\ref{WGC}) in flat space-time, as we described above: the black hole decay must result into states which can not form black-holes, because the latter would exhibit a naked singularity and are therefore forbidden by the Weak Cosmic Censorship. 

The paper is organised as follows. In Section 2, we present the metric describing Reissner-Nordstr\o m black holes in de Sitter space and give our conventions. The limiting case of charged black holes in flat space-time  and neutral black holes in de Sitter space are briefly reviewed in Section 3. This  allows to understand these limiting cases and grab some features of the more general case that we study in Section 4. The form of the dS-WGC is then obtained and discussed in some details. In Section 5, we study the example of gauged R-symmetry in $N=1$ supergravity. Finally, Section 6 contains our conclusions and outlook.

There is nowadays an enormous literature discussing different aspects of the black hole solutions considered here. Thus, we expect that many, probably most, of our intermediate formulae and equations have appeared in other places, with mainly an overlap with the pioneering work of \cite{Romans:1991nq}. As the spirit of this work is different and may be relevant to a different set of readers, we have chosen to be self-contained, therefore reproducing all relevant calculations.

\section{Reissner-Nordstr\o m  charged black hole in de Sitter space-time}

We consider the four-dimensional Einstein--Maxwell  action
\begin{equation}
S= \int d^4x \sqrt{- g}  \left[\frac{1}{16 \pi G} (R -2\Lambda)  +\frac{1}{4 g^2} F^{\mu \nu} F_{\mu \nu} \right]
\label{action}
\end{equation}
where $\Lambda$ is the cosmological constant, $F^{\mu \nu}$ is the electromagnetic field strength tensor with $g$ the $U(1)$ gauge coupling, $R$ is the Ricci scalar and $G$ is the Newton constant. 
A  solution to the equations of motion of (\ref{action}) is given by the metric describing a Reissner-Nordstr\o m  charged black hole \cite{Kottler}. It takes the form:
\begin{equation}
ds^2= -f(r) dt^2 +\frac{dr^2}{f(r)} + r^2 (d\theta^2 + \sin^2 \theta d\phi^2)
\label{metric}
\end{equation}
 in usual Schwarzschild coordinates $(t,r,\theta,\phi)$. We have chosen to work with a space-time signature $(-,+,+,+)$ and a global coordinate system that allows us to connect to the asymptotically flat space-time limit (at large $r$).  In natural units (${\hbar } =c =1$), the metric function $f(r)$ is given by:
\begin{equation}
f(r)=1 -\frac{2 G m}{r} + \frac{G g^2 { q}^2}{4 \pi r^2} - \frac{\Lambda}{3} r^2\,,
\label{metric For Time}
\end{equation}
where $g q $ is the physical black hole charge and $m$ its mass. 

It is convenient to 
use the quantities $M, Q, l$, all having dimension of length:
\begin{eqnarray}
 G m = \frac{  \kappa^{2} m}{8 \pi }  = M \quad,\quad \frac {G g^2 { q}^2} {4 \pi  }=  \frac {\kappa^{2} g^2 { q}^2} {32 \pi^{2} } =  Q^2 \quad,\quad  \Lambda = {\frac {3}{l^2}}\,,
\label{MQLdef}
\end{eqnarray}
where $\kappa^{2}={8 \pi G}$ is the gravitational coupling,
giving:
\begin{equation}
f(r)=1 -\frac{2  M}{r} + \frac{ Q^2}{ r^2} - \frac{r^2}{l^2} \,.
\label{metricf G=1}
\end{equation}
It is useful to note that
\begin{equation}
\frac{M^2}{Q^2}=\frac{\kappa^{2}}{2}\frac{m^2}{g^2 q^2}
\label{MQratio}
\end{equation}
is the ratio constrained by the WGC.

It is straightforward to generalise the above metric to the case with both electric $Q$ and magnetic $Q_m$ charges through the substitution $Q^2 \rightarrow Q^2 +Q^2_m$ with
\begin{eqnarray}
 Q^2_m = q_m^2 g_m^2 \quad,\quad  g_m^2 =\frac {4 \pi^2  } {g^2 } \,,
 \label{magnetic-charge}
\end{eqnarray}
where $g_m$ is the gauge coupling in the dual magnetic theory and $q_m$ is the corresponding charge.

\section{Flat $\Lambda =0$ and neutral $Q =0$ limits}

Let us start with a brief review of the simple limiting cases of $\Lambda =0$ and $Q\neq 0$, or $Q=0$ and $\Lambda\neq 0$, which will make easier to the reader to follow our study of the more general case in the next section.

In the case of flat space time, $\Lambda =0$, the horizon is given by solving
\begin{equation}
 f(r)=1 -\frac{2  M}{r} + \frac{Q^2}{ r^2} =0\,,
\label{metricf L=0}
\end{equation}
which gives an (inner) Cauchy horizon $r_-$ and an (outer) event horizon $r_+$:
\begin{eqnarray}
r_- =  M \left(1-\sqrt{1-\frac { Q^2} {  M^2} }\right) \quad,\quad r_+=M \left(1+\sqrt{1-\frac { Q^2} { M^2} }\right)\,.
\label{equationFlat}
\end{eqnarray}
It is useful to identify, in the process of solving the Einstein equations, how the different terms in $f(r)$ arise from the respective contributions to the energy-momentum tensor. The $Q^2$ term is due to the self-repulsive electromagnetic energy density, while the $-2M$ is due to the  black hole mass attractive force. Different cases can be studied when balancing against each other these two terms. From this point of view, in the $r,t$ coordinates, space-time can be divided in different regions depending on which contribution dominates. We thus have three cases:
\begin{itemize}
\item $Q^2 < M^2$, the two roots of \eqref{metricf L=0} are real and positive, with $0 < r_- < r_+$.  The inner horizon, corresponding to the surface $r=r_-$, coincides with the Cauchy horizon and hides the singularity at $r=0$ from observers in the rest of space-time. In the region $0 < r < r_-$, the coordinate $t$ is time-like, while in the interior of the black hole, corresponding to the region  $r_-  < r <  r_+ $, $t$ is space-like. Finally, in the region $r> r_+$, outside the event horizon at $r=r_+$, the coordinate $t$ is time-like again.

\item $Q^2 = M^2$, the two horizons are degenerate. This corresponds to the extremal black hole case, where the gravitational attraction and electric repulsion are balanced.

\item $Q^2 > M^2$,  the two roots are complex, there is no horizon and the metric (\ref{metric}) exhibits a naked singularity at the origin $r=0$.

\textit{The repulsive electric force became stronger than gravity and forbids the presence of black hole horizons.}
\end{itemize}

In asymptotically flat space-time, the kinematics require the existence of states with $Q ^2 > M^2$ for extremal charged black holes to decay. These super-extremal states can not collapse to form new black holes as this would lead to configurations, with naked singularities, forbidden by the Weak Cosmic Censorship. Insuring that states with $Q ^2 > M^2$ exist allows therefore to avoid the presence of remnants when black hole decay. This is the logic behind the requirement of gravity to be ``the weakest force" that we will generalise in the next section to the case of positive vacuum energy.

In the case of de Sitter space-time, $\Lambda \neq0$ but $Q =0$ (Schwarzschild de Sitter black hole), the horizon is given by considering the zeros of the polynomial
\begin{equation}
f(r)=1 -\frac{2  M}{r}  - \frac{r^2}{l^2}  \, ,
\label{metricf Q=0}
\end{equation}
thus solving:
\begin{equation}
P_0(r)\equiv -l^2rf(r) =  {r^3} -{r}{l^2}+ 2 {M}{l^2}\equiv (r- r_{- -})(r-r_+)(r-r_C) =0\,.
 \end{equation}

Obviously, since the product of roots is $-2Ml^2$, one of them is negative, that we choose to denote as $r_{- -}$, and plays no role in the physics. Moreover, since the sum of the roots vanishes and one is negative, the other two (denoted as $r_+$ and $r_C$) are either both real and positive (with $r_+\le r_C$) or complex. When they are both real, $r_+$ and $r_C$ correspond to the event and cosmological horizons, respectively.
The solutions of the above equation can be written as: 
\begin{eqnarray}
 r_{- -} &= & - \frac{ e^{i \frac{ \pi }{3} } \, \,  l^{4/3}} {\sqrt{3}  \,  ( \,  { - \sqrt{27} M + \sqrt{27 M^2-l^2}}  \,  )^{1/3}}  -  \frac {{e^{- i \frac{ \pi }{3}} } \, \, l^{2/3}  \, \,   ( \,  { - \sqrt{27} M + \sqrt{27 M^2-l^2}}  \,  )^{1/3}}{3^{2/3}}
 \\
 r_+ &= & - \frac{ e^{- i \frac{ \pi }{3} }\, \,   l^{4/3}} {\sqrt{3}  \,  ( \,  { - \sqrt{27} M + \sqrt{27 M^2-l^2}}  \,  )^{1/3}}  -  \frac {{e^{ i \frac{ \pi }{3}} } \, \,  l^{2/3}  \, \,   ( \,  { - \sqrt{27} M + \sqrt{27 M^2-l^2}}  \,  )^{1/3}}{3^{2/3}}
 \\
 r_C &= & \frac{l^{4/3}} {\sqrt{3}  \,  ( \,  { - \sqrt{27} M + \sqrt{27 M^2-l^2}}  \,  )^{1/3}}+\frac{l^{2/3}   \, \,  ( \,  { - \sqrt{27} M + \sqrt{27 M^2-l^2}}  \,  )^{1/3}}{\sqrt{3}}\,.
\label{solution}
\end{eqnarray}
\begin{figure}[ht]
\centering
\includegraphics[scale=0.5]{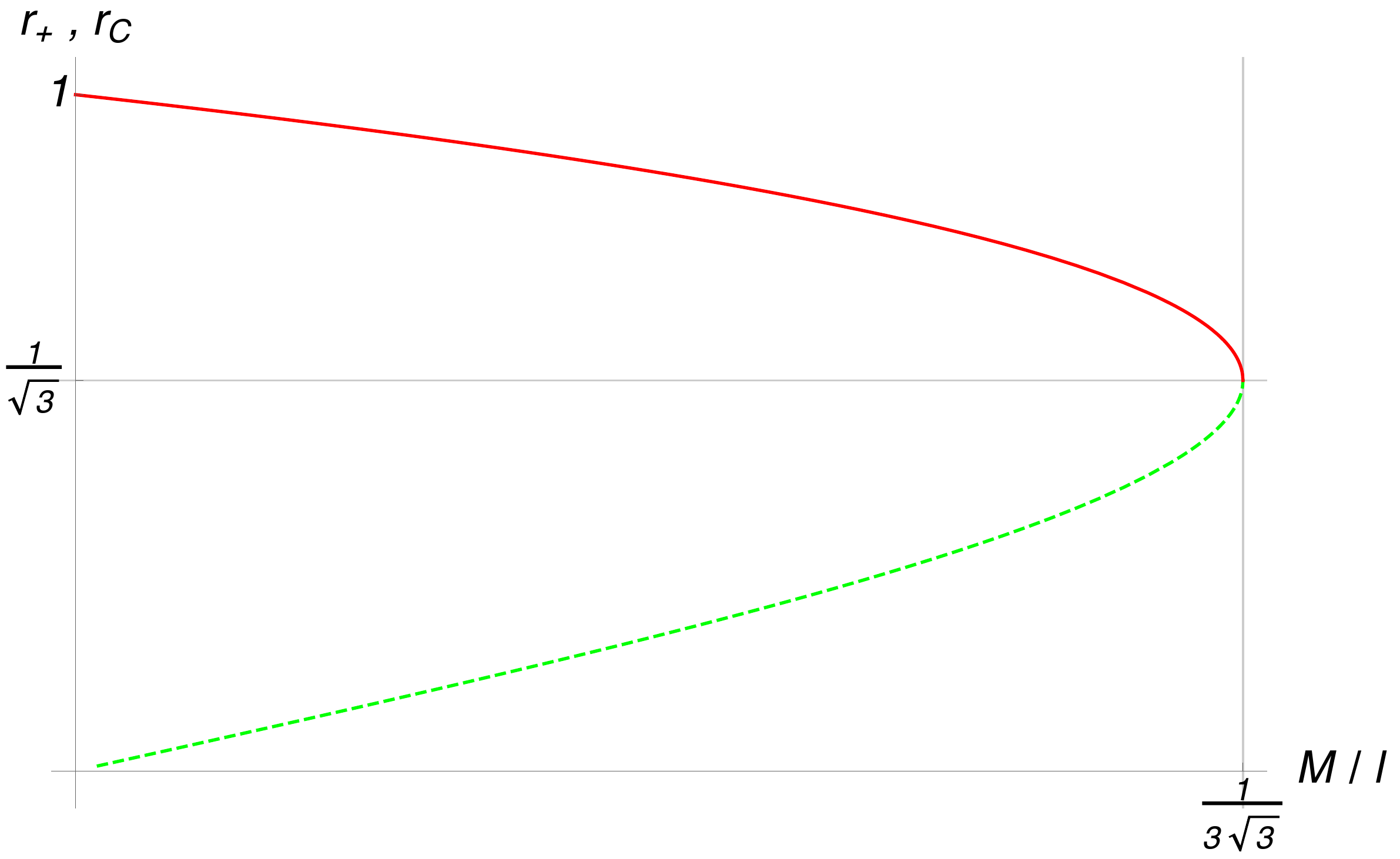}
\captionsetup{width=0.85\textwidth}
\captionsetup{font=small}
\caption{The positive roots of $P_0(r)$: the red curve corresponds to the cosmological horizon and the dashed green one to the event horizon. As $M$ approaches its maximal allowed value $M = \frac {l}{\sqrt{27}}$,  the two roots coincide. This corresponds to a neutral Nariai black hole. }
\label{Q2=0M2}
\end{figure} 

For the purpose of this work, we are interested in their behaviour when varying the parameter $M^2/l^2$, that will be generalised below for charged black holes.  We have again three different regions:
\begin{itemize}
\item $M^2 < \frac{l^2}{27}$, there are two positive roots $0 < r_+< r_C$, where $r_+$ is the black hole event horizon and $r_C$ is the cosmological horizon. The coordinate $t$ is space-like in the region $0 < r< r_+$, and in the region $r > r_C$. \item $M^2 = \frac{l^2}{27}$, the two horizons are degenerate. This corresponds to an extremal case. When the two horizons approach each other, the Schwarzschild coordinates are no more appropriate, one has a (neutral) Nariai black hole and more appropriate coordinates need to be introduced \cite{Ginsparg:1982rs}.
\item $M^2 > \frac{l^2}{27}$,  the two roots are complex, $f(r)$ is always negative, there is no horizon and the metric presents a naked singularity at $r=0$. It is a big space-time crunch or, following \cite{Ginsparg:1982rs}, it represents a de Sitter space eaten by a giant black hole.
\end{itemize}
For illustration, we plot in Fig.~\ref{Q2=0M2} the event and cosmological horizons as a function of $M/l=mG/l$ in the region where they exist, $M^2\le l^2/27$.

\section{Charged $Q \neq 0$ black hole in de Sitter $\Lambda \neq 0$: }

Let us consider now the case of de Sitter $ l^2 >0$ and look for the loci of the horizons. They are obtained by solving
\begin{equation}
f(r)=1 -\frac{2  M}{r} + \frac{ Q^2}{ r^2} - \frac{r^2}{l^2} =0\,,
\label{equationde Sitter}
\end{equation}
which has four solutions $r_{- -}$, $r_-$, $r_+$ and $r_C$ whose nature and values depend on the parameters $M,Q,l$ in (\ref{equationde Sitter}). These are roots of the quartic polynomial:
\begin{equation}
P(r)\equiv -r^2 f(r) = {l^{-2}} r^4 -r^2 +{2  M} r - {Q^2}\,.
\label{Polyde Sitter}
\end{equation}
Since the product of the four roots of the polynomial $P(r)$ is given by $-Q^2<0$, $P$ has necessarily two real roots, one positive and one negative. The remaining two roots are either complex conjugate, or real of the same sign, depending on the sign of its discriminant $\Delta$, negative or positive, respectively. The discriminant $\Delta$ is given by:
\begin{equation}
\Delta = \frac{16}{l^6}{ \left(- \frac{27 }{l^2} (M  l)^4+ \left(l^2+36 Q^2\right) (M l)^2 - Q^2 \left(l^2+4 Q^2\right)^2 \right)} \,.
\label{resultDelta}
\end{equation}
We have therefore three cases:
\begin{enumerate}
\item ${\Delta} > 0$, then $r_{- -}<0$ is not physical and $0< r_-< r_+< r_C$. The  inner and outer horizons $r_-$ and  $r_+$ are, respectively, the locations of the Cauchy and event horizons of the Reissner-Nordstr\o m black hole. The $r=r_C$ surface is the cosmological horizon of the de Sitter space, it satisfies the limit $r_C \rightarrow \infty$ as $\Lambda \rightarrow 0$ where we recover the Reissner-Nordstr\o m case discussed above, in Section~3.
\item ${\Delta} = 0$, then $r_{- -}<0$ and either $0< r_- = r_+<r_C$, or $0<r_-<r_+ = r_C$. Two horizons coincide; the inner and outer, or the outer and cosmological.
\item ${\Delta} < 0$, then $r_{- -}<0$ and $r_C >0$ but $r_- = r_+^*$ are complex and conjugate (or $r_->0$ and $r_C = r_+^*$ complex). There is only one horizon.
\end{enumerate}

The discriminant $\Delta$  is itself a quadratic polynomial of $M^2 l^2$ with discriminant:
\begin{equation}
\delta = \frac{256 \left(l^2-12 Q^2\right)^3}{l^{14}}
\label{defDelta}
\end{equation}
The roots of $\Delta$ are then given by:
\begin{eqnarray}
M^2_{-} (l,Q^2)  &=& \displaystyle{\frac{1}{54 l} 
\left[ {l(l^2+ 36  Q^2) - \left(l^2-12 Q^2\right)^{3/2}} \right]}\,, \nonumber \\
 M^2_{+} (l,Q^2) &=& \displaystyle{\frac{1}{54 l} 
 \left[ {l(l^2+ 36  Q^2) + \left(l^2-12 Q^2\right)^{3/2}} \right]}\,, 
\label{Mmoinsplus}
\end{eqnarray}
 and $\Delta$ is negative outside these roots (for $l^2\le 12Q^2$).
 
 As $r_{- -}$ is always negative, we focus on  $r_-$, $r_+$ and $r_c$.  They define four regions of space:
\begin{itemize}
\item \textbf{Region IV: $  r > r_C$}. In this region, the coordinate $t$ is space-like ($f(r) <0 $). The term ${r^2}/{l^2}$ is dominant, $f(r) \sim -r^2$. Thus, the cosmological constant dominates over the attractive gravitational potential due to mass $M$. The electromagnetic repulsive potential  ${g^2 Q^2}/{r^2}$ is also subdominant. 
\item \textbf{Region III: $ r_+ \le r  \le r_C$}. In this region, the coordinate $t$ is time-like ($f(r) >0 $). The constant term in $f(r) \sim 1 + \cdots $ dominates.
\item \textbf{Region II: $  r_- \le r  \le r_+$}. The coordinate $t$ is space-like again ($f(r) <0 $). The attractive gravitational contribution, corresponding to the term $-2 M / r$, is the dominant one in this region. This is the interior of the black hole, where $r_+$  and $r_-$ are the event and the Cauchy horizons, respectively.
\item \textbf{Region I: $  0  < r  \le  r_-  $} (thus $f(r) >0 $). In this region, the repulsive  electromagnetic force due to the charge $Q$, corresponding to the term $Q^2 / r^2$, dominates the attractive gravitational contribution due to the black hole mass $M$. 
\end{itemize}

For $Q=0$, the inner horizon vanishes $r_-=0$ and we retrieve the discussion in the previous section. $M_-(l,0)$ in \eqref{Mmoinsplus} vanishes and there is no minimal bound for the mass.
As $M$ increases, the attractive potential is more important, the event horizon $r_+$ increases till it reaches the cosmological horizon (see Fig.~\ref{Q2=0M2}). This happens at the maximal value $M=M_{max} (l,0)$, 
\begin{equation}
M^2_{max}(l,0)= M^2_{+} (l,0) = \frac{1}{27 } l^2 \, .
\end{equation}
As it approaches $r_+ \simeq r_C$, we have a neutral Nariai black hole. For bigger masses, there is a naked singularity.

\begin{figure}[ht]
\centering
\includegraphics[scale=0.6]{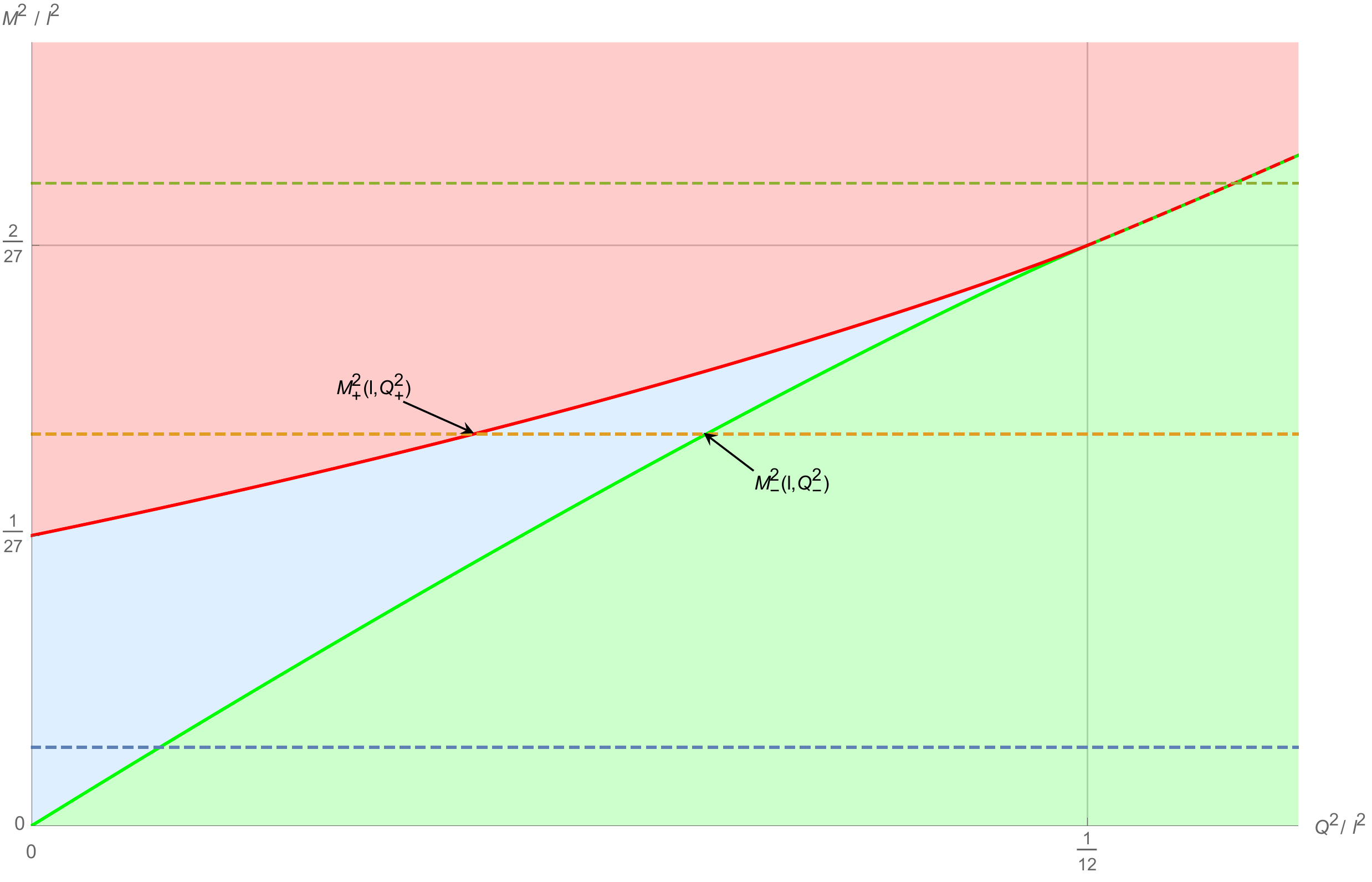}
\captionsetup{width=0.85\textwidth}
\captionsetup{font=small}
\caption{The blue filled region corresponds to the range of $M^2/l^2$ and $Q^2/l^2$ allowing the black hole to possessing three horizons. The solid red curve corresponds to the maximal mass for given charge, $M^2 =M^2_{+} (l,Q_{+}) $, where two roots are degenerate $r_+=r_C$. The solid green curve represents the minimal mass $M^2 =M^2_{-} (l,Q_{-}) $, where $r_- =r_+$ describing cold extremal black holes. The green filled area corresponds to the region allowed by the de Sitter Weak Gravity Conjecture (dS-WGC) for the (mass, charge) parameters.}
\label{Q2M2}
\end{figure}

Let us now fix the black hole mass $M$ and turn on the electric charge. Or, one might of course fix instead $Q \neq 0$ and vary $M$ for the above discussion. As $Q$ increases,\footnote{For convenience, we choose $Q$ positive.} the ``sizes" of the different regions change depending on the balance between the different repulsive and attractive interactions. 
Given the mass $M$, it is useful to define $Q_{\pm}$ as the values of the charge that satisfy:
\begin{eqnarray}
M^2_{-} (l,Q_-^2)  = M^2 \quad, \quad
M^2_{+} (l,Q_+^{2}) =M^2\,,
\label{Qmoinsplus}
\end{eqnarray}
where the functions $M_\pm$ are given in \eqref{Mmoinsplus}.
To illustrate the discussion below, we we plot in Fig.~\ref{Q2M2} the curves of $M^2_\pm$ in the plane (mass, charge) of the black hole ($M^2/l^2$ vertical axis, versus $Q^2/l^2$ horizontal axis). 

We have now the following different cases ($M$ has fixed value):
\begin{enumerate}

\item  $M^2 < \displaystyle\frac{l^2}{27}$: here $Q_+$ does not exist because $Q_+^2$ is negative (see Fig.~\ref{Q2M2} and follow the lower horizontal blue line).
\vspace{0.2cm}
\begin{itemize}
\item  As $Q \nearrow$, $Q <Q_-$, we have $r_- \nearrow$,  $r_+ \searrow$ and $r_C \nearrow$, while $M>M_-(l,Q^2)$. The region II shrinks with $r_{+} \rightarrow r_{-}$.  Note that it corresponds to the interior of the black hole between the Cauchy surface and the event horizon. As the charge increases further, a critical value is reached when $Q=Q_{-}$. At this value, the black hole has its two horizons degenerate $r_- = r_+$ {\footnote{ More precisely, approaching this region the $(t,r)$ coordinates are no more appropriate and after such a coordinate change, one can study the resulting charged Nariai solution. The need of a new coordinate system remains a priori true for all cases where two roots collide.}}.  
\vspace{0.3cm}
\item Beyond this value, i.e. $Q> Q_{-} $ and $M^2 < M^2_{-} (l,Q^2) $, we have $\Delta < 0$, so there is only one (positive) root for the polynomial $P(r)$ in~\eqref{Polyde Sitter}. The region II has disappeared and the regions I and III have merged. We have two regions (I+II+III) and (IV) separated by a cosmological horizon. The region I has the coordinate $t$ time-like with a naked singularity.  The other region (IV) has $t$ space-like.
\vspace{0.3cm}

\textit{The repulsive electrical energy density has become too strong and forbids the presence of the black hole horizons.}

\end{itemize}

\vspace{0.3cm}
\item $\displaystyle\frac{l^2}{27} \leq M^2 \leq \frac{2 l^2}{27}$: now both $Q_\pm$ exist with $Q_+<Q_-$ and there is a region between $\left[Q_+,Q_-\right]$ where $M \in \left[M_-(l,Q),M_+(l,Q)\right]$ such that we have three horizons (see Fig.~\ref{Q2M2} and follow the red middle horizontal line within the blue area).
Varying $Q$ towards leaving this region, we have two cases:
\vspace{0.2cm}
\begin{itemize}
\item [$\bullet$] Decreasing $Q$, $Q \rightarrow Q_+$: When $Q=Q_+$, two roots of $P(r)$ are degenerate $r_+ = r_C$. As  $Q \searrow$ we have $r_- \searrow$,  $r_+\nearrow$ and $r_C \searrow$. The region III shrinks until it disappears for $Q<Q+$. In this range of charge, $0< Q< Q_+$ (in the red area of Fig.~\ref{Q2M2}), we have only one horizon $r_-$, separating two regions: (I) where the coordinate $t$ is time-like and the merged (II+III+IV) region where $t$ is space-like. 
\vspace{0.3cm}

\textit{The repulsive electrical energy density is too weak to forbid the black hole from eating the de Sitter space-time.}
\vspace{0.3cm}

\item Increasing $Q$, $Q \rightarrow Q_-$: As $Q \nearrow$, we have $r_- \nearrow$,  $r_+\searrow$ and $r_C \nearrow$. For $Q=Q_-$, two roots of $P(r)$ are degenerate $r_+ = r_-$ and beyond this point, for $Q >Q_-$ (in the green area of Fig.~\ref{Q2M2}), the region II disappears and there is only one horizon $r_C$ separating again two regions: (I+II+III) where $t$ is time-like and  (IV) where $t$ is space-like.
\vspace{0.3cm}

\textit{The repulsive electrical energy density has become strong enough to forbid the presence of the black hole horizons.}

\end{itemize}

\vspace{0.3cm}

\item  $M^2 = \displaystyle\frac{2 l^2}{27}$: for $Q^2 \rightarrow l^2/12$, then $r_- \rightarrow r_+ \rightarrow r_C \rightarrow l/{\sqrt{6}}$, the three roots of $P(r)$ become degenerate when the limits for $M,Q$ are reached.  At this point, one has again a Nariai black hole and our $(t,r)$ coordinate system fails to grasp the physics \cite{Ginsparg:1982rs}. However, one can qualitatively argue that at this point the added repulsive effects of the electric charge and cosmological constant are just at the verge of forbidding the attractive gravity, due to the mass, from forming a black hole event horizon for $Q>l/\sqrt{12}$ (see Fig.~\ref{Q2M2}).

\vspace{0.3cm}

\item $M^2 > \displaystyle\frac{2 l^2}{27}$: there is only one horizon. Its location moves as a function of $(Q,M)$, away from the point discussed above where the three horizons coincide (see Fig.~\ref{Q2M2} and follow the upper horizontal green line). Let us introduce the parametrisation:

\begin{equation}
M= {\sqrt{\frac{2 }{27}}}\,  l + \delta M \, , \qquad Q^2=  \frac{l^2}{12}+ \sqrt{\frac{2}{3}} {\delta Q^2}
\end{equation}

For $ \delta M > {\delta Q^2}/ l$, there is a continuation of the solution we interpreted as a black hole that eats the de Sitter space (red area in Fig.~\ref{Q2M2}). For $ \delta M < {\delta Q^2}/ l$, there is a space-time filled by the electromagnetic energy density and a cosmological horizon. In this case, the electromagnetic self-repulsion forbids again the presence of the black hole event horizon (green area in Fig.~\ref{Q2M2}). For $ \delta M = {\delta Q^2}/ l$, the forces are balanced.
\end{enumerate}

\vspace{0.2cm}

Requiring the repulsive interaction to be strong enough to forbid the appearance of black hole horizons, we are led to the following conditions for the dS-WGC imposing the existence of a state in the green area of Fig.~\ref{Q2M2}:
\begin{enumerate}
\vspace{0.2cm}
\item  \underline{\textbf{Small charge:}} \, \,   $Q^2\leq \displaystyle{l^2\over 12}$ $\left(g^2 q^2 \leq \displaystyle{\pi l^2\over 3G}\right)\quad$ and 
$\quad M^2 \leq \displaystyle\frac{2}{27}  l^2$   $\left(m^2 \leq \displaystyle{ \frac{2}{27 G^2 } l^2}\right)$:

\begin{eqnarray}
M^2 \!\!\!\!\!&<&\!\!\!\!\! M^2_{-} (l,Q^2)  = \frac{1}{54 l} \left[ {l(l^2+ 36  Q^2) - \left(l^2-12 Q^2\right)^{3/2}} \right]
\quad \nonumber\\[5pt]
 &\iff& \quad m^2<\displaystyle{\frac{1}{54 G^2 l} \left[ {\frac{9 G l }{\pi } g^2
   q^2 +l^3 - \left(l^2-\frac{3 G }{\pi } g^2 q^2 \right)^{3/2}} \right] }
\label{condition-light}
\end{eqnarray}

\item \underline{\textbf{Large charge:}} \, \,  $Q^2 \geq \displaystyle{l^2 \over 12}$ $ \left( g^2 q^2 \geq \displaystyle{\pi l^2\over 3G}\right)$:
\begin{equation}
M^2<\frac{3}{2} \frac{1}{l^2}\left(Q^2+{5\over 36} l^2 \right)^2
\quad    \iff \quad   
m < \frac{5 }{12 \sqrt{6} } {l\over G} + \sqrt{\frac{3}{2}} {g^2q^2\over 4\pi l} \, . 
\label{condition-heavy}
\end{equation}
\end{enumerate}

\subsection{Small and large curvature limits of the dS-WGC:}

In the flat space limit $l\to\infty$, the second region \eqref{condition-heavy} disappear, while the first region leads to
\begin{equation}
\label{flatlimit}
M^2 <Q^2 -\frac{Q^4}{l^2}-2 \frac{Q^6}{l^4}+ {\cal O}(1/l^6) 
 \quad \iff \quad
m^2 < \frac{g^2 q^2}{4 \pi  G}-\frac{g^4 q^4}{16 \pi ^2 l^2}-\frac{G g^6  q^6}{32 \pi ^3
   l^4}+O\left(\frac{1}{l^6}\right)
\end{equation}
reproducing the known WGC in flat space
\begin{equation}
\label{flatlimit1}
gq > \sqrt{4\pi G}\, m\qquad (l\to\infty)\,,
\end{equation}
while in the presence of a small positive cosmological constant there is a correction; it implies that the minimal charge of the required state becomes:
\begin{equation}
\label{flatlimit2}
Q^2 > M^2 +\frac{M^4}{l^2}+ {\cal O}(1/l^4) \quad\Rightarrow\quad
g q >  \sqrt{4\pi G}  m \left(1+{G^2 m^2\over 2l^2}+\cdots\right)\,.
\end{equation}

In the opposite limit of strong curvature, $l\to 0$, the first region \eqref{condition-light} disappears and the second  region leads to
\begin{equation}
Q^2 > \sqrt{2\over 3} lM - {5\over 36}l^2 \quad\Rightarrow\quad Q > \left({2 \over 3}\right)^{1/4}   \sqrt{lM}\left(1+{\cal O}(l)\right). 
\label{highcurvlimit}
\end{equation}
Note that this limit is independent of the Newton constant at leading order:
\begin{equation}
\label{highcurvlimit1}
gq > \left({32 \pi^2 \over 3}\right)^{1/4} \sqrt{lm}\qquad (l\to 0) \,,
\end{equation}
since the factors of $G$ drop, as opposed to \eqref{flatlimit1}.

\subsection{Magnetic black hole in de Sitter}

Another condition leading to a form of the WGC in de Sitter space was proposed in \cite{Huang:2006hc}.  For a $U(1)$ gauge theory, the size of the minimally charged monopole is required to lie in between the event and cosmological horizons. The size of this monopole is of order $1/\Lambda_m$,  with $\Lambda_m$ the cut-off of the theory, while its mass is $\Lambda_m/g$ with $g$ the gauge coupling. The magnetic Reissner-Nordstr\o m black hole that would be created by such a monopole has as metric (\ref{metric}) with:

\begin{equation}
f(r)=1 -\frac{2  \Lambda_m}{g^2 r} + \frac{ q_m^2}{ g^2 r^2} - \frac{r^2}{l^2} \,,
\label{metricf Magnetic}
\end{equation}
where $q_m$ is the magnetic charge (see \eqref{magnetic-charge}). In \cite{Huang:2006hc}, the term due to the monopole charge was left out which amounts to put $q_m=0$ in our formulae. The condition that the size of the monopole lies between the event and cosmological horizons implies:
\begin{equation}
f(\frac{1}{\Lambda_m})=1 + \frac{ \Lambda_m^2}{ g^2 }( q_m^2 - 2)  - \frac{1}{\Lambda_m^2 l^2} \ge 0\,,
\label{metricf Magnetic2}
\end{equation}
that can be written as:
\begin{equation}
\frac{ ( 2-q_m^2 ) l^2}{ g^2 } \Lambda_m^4 -  l^2 \Lambda_m^2 +1 \le 0 \, .
\label{metricf Magnetic3}
\end{equation}
This inequality can be satisfied if the discriminant of the quadratic polynomial in $\Lambda_m^2$ is positive which amounts to having:
\begin{equation}
g \ge \frac{2  \sqrt{( 2-q_m^2 )} }{ l }  \qquad \xrightarrow[q_m = 1] { }   \qquad  g \ge \frac{ 2}{ l } \, ,
\label{g-monopole-condition}
\end{equation}
where we have considered the magnetic black hole to carry the minimal charge. In the extremal case, $g = \frac{ 2}{ l }$, the size of the monopole reaches the cosmological horizon $1/\Lambda_m = l /\sqrt{2}=r_C$ and the magnetic monopole has a mass $M_m= l/ \sqrt{8} $.

In \cite{Huang:2006hc},  the weak gravity conjecture  condition was identified with (\ref{g-monopole-condition})  but this contains an implicit mass dependence through the de Sitter radius as we have shown. Taking both relations charge and mass with respect to $l$, we find that the conditions (\ref{condition-light}) and  (\ref{condition-heavy}), 
 i.e. absence of a Reissner-Nordstr\o m black hole configuration,  is sufficient to insure that  the condition (\ref{g-monopole-condition}) from the restriction of the monopole size, is also satisfied. Indeed, the values of $(M_m^2,Q_m^2)=(l^2/8,l^2/4)$ fall deep in the region described by the condition (\ref{condition-heavy}).

\section{The $U(1)$ R-symmetry case:}

We now study an example of a $U(1)$ gauge symmetry in de Sitter background in the context of $N=1$ supergravity. To avoid the problem of obtaining dynamically a de Sitter vacuum by minimising the scalar potential, we consider the simplest case of a positive contribution to the cosmological constant due a Fayet-Iliopoulos D-term. This implies that the $U(1)$ is an abelian gauge R-symmetry $U(1)_R$~\cite{FVP}, which remains unbroken if a superpotential is absent or vanishes at the minimum. The minimal model is pure supergravity coupled to a single vector multiplet that gauges the R-symmetry~\cite{Freedman}. 
It was recently noticed that this model is dual under electromagnetic duality to a supergravity theory with deformed supersymmetry, or equivalently to a magnetic Fayet-Iliopoulos term~\cite{Antoniadis:2020qoj}.

Since the gauging of the R-symmetry induces both a cosmological constant $\Lambda$ and a charge $q_R$ of the fermions (gravitino and gaugino), these are not independent but are related in terms of the Newton's constant. Setting for simplicity here the gravitational coupling to unity, $\kappa=1$, one finds the relation~\cite{Antoniadis:2020qoj}:
\begin{equation}
\Lambda=2q_{3/2}^2\quad\Rightarrow\quad q_{3/2}^2={3\over 2l^2} \quad \,{\rm and}\, \quad Q_{3/2}^2={3\over 64\pi^2l^2} 
\,,
\label{qccR}
\end{equation}
where $q_{3/2}$ is the physical R-charge of the gravitino, including the gauge coupling: $q_{3/2}=qg$. On the other hand, the gravitino (and gaugino) mass term in the Lagrangian vanishes since the superpotential is zero.

The value \eqref{qccR} corresponds  to 
 \begin{equation}
 \frac {Q_{3/2}^2}{l^2} = \frac{3}{64\pi^2 l^4} \quad \, \Rightarrow \,  \quad \bigg\{   \frac {Q_{3/2}^2}{l^2}  \xrightarrow[l \rightarrow + \infty] { } 0; \quad \, \,  \frac {Q_{3/2}^2}{l^2}  \xrightarrow[l \rightarrow 0] { }  \infty\bigg\}\,,
 \label{QoverlR}
 \end{equation} 
which spans all positive values on the horizontal axis of Fig.~\ref{Q2M2}. The maximal squared-mass of the gravitino, if it plays the role of the dS-WGC state, is given by 
(\ref{condition-heavy})  and (\ref{condition-light})  for small and large $l$, respectively, with a junction at $l= {\sqrt{3/4 \pi}}$ , i.e. $\Lambda = 4 \pi$, $Q_{3/2}^2 = l^2/12= 1/16 \pi$ and $M_{3/2}^2= {1}/{18 \pi}$.  
\begin{itemize}
\item For $ l \in  \left. \right] 0,{\sqrt{3/4 \pi}} \left. \right] $, the dS-WGC condition (\ref{condition-heavy})  requires a maximal squared-mass which goes as $1/l^6$ for small $l$ and spans the interval $M_{3/2}^2 \in  \left. \right[ {1}/{18 \pi}, + \infty  \left. \right[ $. 
\item For $ l \in  \left. \right[  {\sqrt{3}/\sqrt{4 \pi}}, + \infty  \left. \right[ $,  the dS-WGC condition (\ref{condition-light})  requires a maximal squared-mass $M_{3/2}^2=M^2_{{3/2}-} (l,Q_{3/2}^2) \in  \left. \right] 0, {1}/{18 \pi} \left. \right] $ which behaves as $1/l^2$ for large $l$. This behaviour is reminiscent of the $m^2_{3/2} M^2_{Pl}$ contribution of the gravitino mass to the vacuum energy in $N=1$ supergravity.
\end{itemize}
On the other  hand, there is only one possible gravitino mass, constant or with a universal power law dependence for all values of the de Sitter radii, which is compatible with the dS-WGC: $m_{3/2}=0$.

Note that the above value of the R-charge in \eqref{qccR} is dual to a minimal magnetic charge ${\tilde q}_{3/2}=2\pi/q_{3/2}$, for which ${\tilde Q}_{3/2}^2/l^2=1/(8q_{3/2}^2l^2)=1/12$, the special limiting value in Fig.~\ref{Q2M2}!
It is not clear to us whether there is any deep reason besides this astonishing coincidence.

We can contemplate the possibility that this magnetic charge corresponds to a monopole configuration with a mass $\Lambda_m/g$ that satisfies the dS-WGC. Then, we have:
 \begin{equation}
\frac{\Lambda_m}{g^2} \le \sqrt{\frac{2 }{27}} l \quad \,{\rm and}\, \quad \frac{1}{g^2} = \frac{ 2}{3}  l^2  \quad  \Rightarrow  \Lambda_m \le l/{\sqrt{6}} = r_C\,,
\label{qcc2R}
\end{equation}
where $r_C$ is the cosmological horizon in the magnetic theory. Putting back explicitly the dependence in Planck mass, $M_{Pl} = G^{-1/2}$, this leads to:
\begin{equation}
{\Lambda_m} \le (r_C M_{Pl})\, \,  M_{Pl}   \quad  \Leftrightarrow  \quad  {\Lambda_m} \le \frac{M_{Pl} }{\sqrt{{2 \Lambda}}}M_{Pl}\,.  
\label{qcc3R}
\end{equation}

At a next step, one may introduce charged chiral multiplets. The scalar masses would receive a D-term contribution proportional to the R-charge but there may also be an F-term contribution if an appropriate superpotential is allowed. The D-term potential reads:\footnote{We assume a canonical K\"ahler potential since non-canonical terms involve higher powers of scalar fields that do not change the masses around the origin.}
\begin{equation}
V_D={1\over 2}\left(\sum_i q_i|\phi_i|^2+2q_{3/2}\right)^2\,,
\label{VD}
\end{equation}
where $\phi_i$ are the various scalars with physical R-charges $q_i$. To guarantee that the  gauge symmetry remains unbroken, one should impose the absence of tachyons around the origin for all charged scalars. This constraint implies that all R-charges are positive in the convention where $q_{3/2}$ is positive and given by \eqref{qccR}. The scalar masses $m_i$ of $\phi_i$ are then given by:
\begin{equation}
m_i^2=2q_iq_{3/2}\,,
\label{scalarmasses}
\end{equation}
implying:
\begin{equation}
Q_i^2={q_i^2\over 32\pi^2}\quad;\quad M_i^2={m_i^2\over 64\pi^2}={q_iq_{3/2}\over 32\pi^2} 
\,.
\label{QMi}
\end{equation}

On the other hand, an F-term contribution to the mass may exist if a quadratic superpotential $W$ is allowed. There are two cases: $W={\mu\over 2}\phi_0^2$ or $W=\mu\phi_+\phi_-$ for some fields $\phi_{0,+,-}$, where $\mu$ is a mass parameter. This is possible if $q_0=q_{3/2}$ or $q_++q_-=2q_{3/2}$, respectively, implying:
\begin{eqnarray}
Q_0^2 = Q_{3/2}^2 &=& M_0^2
\label{QM0}
\end{eqnarray}
and
\begin{eqnarray}
Q_+ + Q_- = 2Q_{3/2} &;&  M_\pm^2=Q_\pm Q_{3/2} +\mu^2
\label{QM0+-}
\end{eqnarray}
Obviously, the case (\ref{QM0}) can not be a dS-WGC state. In the case (\ref{QM0+-}), there are two  extrema:
\begin{eqnarray}
 \{Q_\pm = 2Q_{3/2}, \quad   Q_\mp = 0  \} &;&  \{ M_\pm^2=2 Q_{3/2}^2 +\mu^2, \quad  M_\mp^2= \mu^2 \},
\label{QM0+-M}
\end{eqnarray}
which can give a dS-WGC state if 
\begin{eqnarray}
Q_{3/2}^2 \le \frac{1}{32 \pi } &:& l \geq \sqrt{\frac{3}{2 \pi }}  \quad\Rightarrow \quad  \mu^2 <\frac{16 \pi ^3 l^4-2 \sqrt{\frac{\left(4 \pi ^2 l^4-9\right)^3}{l^8}} l^2+27 \pi }{864
   \pi ^3 l^2}\,, \nonumber \\
Q_{3/2}^2 \ge \frac{1}{16 \pi } &:&   l \leq \sqrt{\frac{3}{4 \pi }}  \quad\Rightarrow\quad  \mu^2 <\frac{27}{512 \pi ^4 l^6}-\frac{1}{64 \pi ^2 l^2}+\frac{25 l^2}{864} \, ,
\label{CondQM0+-}
\end{eqnarray}
corresponding to the two conditions (\ref{condition-light}) and (\ref{condition-heavy}), respectively. The fermionic partners of these scalars have a common mass $\mu$ and a charge $Q_\pm - Q_{3/2}$. The maximally charged fermion state will have the same charge $Q_{3/2}$ as the gravitino. Therefore, assuming it plays the role of the dS-WGC state, one can constrain its mass. Bearing the same charge as the gravitino, the constraints on $\mu$ are the same as those we discussed above for the gravitino mass, below \eqref{QoverlR}.

\section{Conclusions}

On the one hand, the swampland program aim is to provide a set of selection rules obeyed by any quantum field theory that can be coupled to quantum gravity. The best established rule is given by the weak gravity conjecture. On the other hand, one can not overestimate the importance of investigations of space-times with non-vanishing vacuum energy as our Universe falls in this class. This motivates the study performed here for the modification of the Weak Gravity Conjecture (WGC) by positive cosmological constant in de Sitter space time.

Our investigation rests on the deep relation between the WGC and the Weak Cosmic Censorship (WCC). This is already the case in Minkowski space-time \cite{ArkaniHamed:2006dz} and in the AdS space. For instance, in the latter case, \cite{Crisford:2017zpi} noted that some space-time configurations would present a naked singularity unless WGC states, with big enough charge-to-mass ratio where the conformal dimension of operators plays the role of mass, are present. Here the logic is reversed, the WCC and absence of black hole remnants imply the dS-WGC. Interestingly enough, in contrast with previous proposals \cite{Huang:2006hc,Montero:2019ekk}, our bound has a smooth limit towards the flat space-time result.

We have discussed the simplest case of de Sitter spacetime obtained in gauged $N=1$ supergravity where the abelian symmetry is the $U(1)_R$ R-symmetry. We have studied that the specific relation between the R-charge and vacuum energy implies that dS-WGC forces the gravitino mass to be smaller than the dS-WGC bound. We have also discussed some features of the magnetic dual theory that may be worth noticing.

Some questions have been left out for future studies. For instance, it is of most importance to test our conjecture on explicit examples. It will also be interesting to study the implications of our bounds in cosmological models.  Let us also mention that one could investigate the nature of the lower bound on masses obtained by \cite{Montero:2019ekk}, its complementarity to the upper bounds obtained here,  the generalisation to other dimensions ($D \neq 4$) and the computation of corrections from higher derivative terms in the action.

\noindent
\section*{Acknowledgments}
The work of K.B. is supported by the Agence Nationale de Recherche under grant ANR-15-CE31-0002 ``HiggsAutomator'', while the work of IA is supported by a CNRS-PICS grant 07964.

\noindent

\providecommand{\href}[2]{#2}\begingroup\raggedright
\endgroup

\end{document}